\documentclass[12pt,preprint]{aastex}
\usepackage{graphicx}
\begin{document}

\title{A New Feature in the Spectrum of the Superluminous LMC Supergiant
HDE~269896}

\author{Mariela A. Corti\altaffilmark{1,2}}
\affil{Instituto Argentino de Radioastronom\'{\i}a (IAR), Centro
Cient\'{\i}fico Tecnol\'ogico de La Plata (CCT--LP), CONICET, C.C. No. 5, 
1894 Villa Elisa, Argentina; and Facultad de Ciencias Astron\'omicas y 
Geof\'{\i}sicas, Universidad Nacional de La Plata, Paseo del Bosque s/n,\\ 
1900 La Plata, Argentina}
\authoremail{mariela@lilen.fcaglp.unlp.edu.ar}

\author{Nolan R. Walborn}
\affil{Space Telescope Science Institute,\altaffilmark{3} 3700 San Martin 
Drive, Baltimore, MD 21218, USA}
\authoremail{walborn@stsci.edu}

\author{Christopher J. Evans}
\affil{UK Astronomy Technology Centre, Royal Observatory Edinburgh,
Blackford Hill, Edinburgh EH9 3HJ, UK}
\authoremail{cje@roe.ac.uk}

\altaffiltext{1}{Visiting Astronomer, Complejo Astron\'omico El
Leoncito (CASLEO), operated under agreement among Consejo Nacional de
Investigaciones Cient\'{\i}ficas y T\'ecnicas (CONICET), Secretar\'{\i}a
de Ciencia y Tecnolog\'{\i}a (SeCyT), and the Universities of 
La Plata, C\'ordoba, and San Juan, Argentina.} 
\altaffiltext{2}{Member of Carrera del Investigador Cient\'{\i}fico, CONICET.} 
\altaffiltext{3}{Operated by the Association of Universities for Research
in Astronomy, Inc., under NASA contract NAS 5-26555.}

\begin{abstract}
We have found strong selective emission of the N~II 5000~\AA\ complex in
the spectrum of the LMC hypergiant HDE~269896, ON9.7~Ia$^+$.
Since this object also has anomalously strong He~II $\lambda$4686
emission for its spectral type, an unusually wide range of ionization in 
its extended atmosphere is indicated.  The published model of this
spectrum does not reproduce these emission features, but we show that
increased nitrogen and helium abundances, together with small changes in
other model parameters, can do so.  The morphological and possible
evolutionary relationships of HDE~269896, as illuminated by the new spectral 
features, to other denizens of the OB Zoo are discussed.  This object may
be in an immediate pre-WNVL (Very Late WN) state, which is in turn
the quiescent state of at least some Luminous Blue Variables.

More generally, the N~II spectrum in HDE~269896 provides a striking
demonstration of the occurrence of two distinctly different kinds of line 
behavior in O-type spectra: normal absorption lines that develop P~Cygni 
profiles at high wind densities, and selective emission lines from the same 
ions that do not.  Further analysis of these features will advance 
understanding of both atomic physics and extreme stellar atmospheres.
\end{abstract}


\section{Introduction}

Inverse N vs.\ C, O anomalies in OB absorption-line spectra, denoted as 
OBN and OBC, have been described by Walborn (1976, 2003).  It is now 
generally accepted that the morphologically normal majority of OB 
supergiants display an admixture of CNO-cycled material in their 
atmospheres and winds, while the relatively rare OBC objects have
physically normal (i.e., main-sequence) CNO abundances, and the OBN 
may have enhanced mixing as a result of additional effects such as binary 
interactions or rapid initial rotational velocities, with homogeneous 
evolution in extreme cases (Maeder \& Meynet 2000). 

For instrumental and historical reasons, as well as sometimes scientific
ones, optical stellar spectroscopy has usually concentrated on certain 
wavelength regions, such as the blue-violet or the red.  As a result, 
new phenomena may be encountered when the orphan regions are examined.     
A recent example is the unexpected discovery of CNO anomalies in O2
spectra, from a survey of the 3400~\AA\ region (Walborn et~al.\
2004; Morrell et~al.\ 2005).  Another example is the subject of this
report, arising from a digital OB spectral-classification atlas extending
somewhat beyond the traditional blue-violet limit around H$\beta$
(http://www.fcaglp.unlp.edu.ar/$\sim$mariela/atlas-mariela/index.htm).   

The spectrum of HDE~269896 (= Radcliffe (R) 129, Sanduleak (Sk)
$-68^{\circ}135$), one of the brightest OB stars in the Large
Magellanic Cloud, was first described in detail by Hyland \&
Bessell (1975) and Walborn (1977).  It displays two unusual
characteristics: an emission line of He~II $\lambda$4686 at its
relatively late spectral type, and the CNO absorption-line intensity 
anomalies that define the ON supergiant class.  At earlier spectral
types, the $\lambda$4686 emission (Of) effect is a luminosity indicator
(Walborn 1971, 1973, 2008; Walborn \& Fitzpatrick 1990); hence, its presence 
in this spectrum was interpreted as an effect of superluminosity, which is
supported by the very bright $M_V$ of $-$8.1, a magnitude brighter than
typical Ia supergiants.  Thus, this spectrum was classified ON9.7~Ia$^+$
by Walborn (1977).

\section{Observations}

The observations for the atlas, including HDE~269896 and several
comparison objects shown below, were obtained by M.A.C. at the CASLEO 
``Jorge Sahade'' 2.15~m telescope in San Juan, Argentina, during
2001--2004.  (HDE~269896 was observed on 2001 February 5.)  The REOSC 
echelle spectrograph (on loan from the Institut d'Astrophysique de Li\`ege) 
was used in its non-echelle mode with a 200--300~$\mu$m slit; a 600~l~mm$^{-1}$ 
grating; and a Tek $1024 \times 1024$, 24~$\mu$m pixel CCD, covering the 
3700--5600~\AA\ range at a reciprocal dispersion of 1.64~\AA~pix$^{-1}$ 
(resolution $\sim3$~\AA) and typical S/N of 150--200.  The exposure
times ranged from 20~s for $\mu$~Normae to 600~s for HDE~269896
($V$~=~11.36, ($B-V$)~=~0.00).  Reductions were performed with standard 
IRAF routines. (IRAF is distributed by NOAO, operated by AURA, Inc., under 
agreement with the NSF.)

We also make use of high-resolution spectroscopy of HDE~269896 from the
Ultraviolet-Visual Echelle Spectrograph (UVES) at the European Southern
Observatory Very Large Telescope (VLT).  These data are described and 
discussed by Evans et~al.\ (2004); they were obtained by L.~Kaper.

\section{Results}

The low-resolution spectrogram of HDE~269896 is shown in Figure~1, together 
those of four Galactic ON, OC, and normal supergiants of the same spectral 
type for comparison.  All of these spectra are also compared by Walborn \&
Fitzpatrick (1990) at twice the resolution, but only to 4750~\AA.
The absorption-line CNO anomalies are best seen here in the ratio of
N~III $\lambda$4640 to C~III $\lambda$4650.  HDE~269896 also stands out
by its strong H$\beta$ emission, weakened H$\gamma$ absorption, and He~II
$\lambda$4686 emission, which are diagnostics of its higher luminosity.
In addition, it is now seen to have a strong N~II 5000~\AA\ emission 
feature, which is present more weakly in the other ON supergiant
HD~105056, but completely absent in the OC and normal objects.  This N~II
feature is a blend of several lines, as shown in high-resolution data for
HD~105056 by Walborn (2001), and more weakly still in some related
northern objects by Walborn \& Howarth (2000).  Thus, it is a further
important diagnostic of nitrogen enhancement, and perhaps luminosity as
well, in late-O supergiants.

Sections of the VLT/UVES observation of HDE~269896 are displayed in
Figures~2, 3, and 4, and compared with three different model fits.  The
model parameters are specified in Table~1, and equivalent-width (EW)
measurements of N~II lines in both the data and the models are listed in
Table~2.  Negative EW values signify emission lines.  The transitions are
shown with the upper levels first, as appropriate for emission lines.
Relative intensities of the lines from the NIST Atomic Spectra Database
(Ralchenko et~al.\ 2008)
are also listed, to emphasize that they are all comparable.  However, as
seen in the figures and table, in the stellar spectrum the lines fall into 
two groups with qualitatively distinct behaviors: those with absorption   
or incipient P~Cygni profiles, and those with narrow, symmetrical
emission profiles.  The latter are typical selective emission lines, as
reviewed and discussed in detail by Walborn (2001).  The UVES data show
that the next N~II downward transitions, in the $\lambda$5700 region, not
discussed here, are also in selective emission in HDE~269896.

\section{Discussion}

The ionization potentials of H$^0$, N$^+$, and He$^+$ are 13.6, 29.6, 
and 54.4~eV, respectively.  It is surprising to see both N~II and He~II
features in the same spectrum; the former corresponds to later and the
latter to earlier spectral types than that of HDE~269896.  Thus, this
superluminous object must have an unusually extended atmosphere with a
large range of ionization conditions.  These features together provide 
constraints on models of this atmosphere, as substantiated below.

Fitzpatrick (1991) discovered an interesting LMC relative of HDE~269896,
namely Sk~$-66^{\circ}169$, O9.7~Ia$^+$.  As implied by the
spectral type, it has similar values of the classification criteria,
including He~II $\lambda$4686 emission (albeit somewhat weaker,
consistent with its fainter but still superluminous $M_V$ of $-$7.5), 
and morphologically normal CNO features.  The optical spectra of
both objects are illustrated by Fitzpatrick (1991), and their UV spectra
from 900 through 1900~\AA\ by Walborn, Parker, \& Nichols (1995) and
Walborn et~al.\ (2002a).  Comparative astrophysical analyses of these two 
stars are of considerable interest. 

Indeed, they have been analyzed similarly by Crowther et~al.\ (2002) and
Evans et~al.\ (2004).  Sections of the spectrum of HDE~269896 analyzed in
the latter paper are reproduced here in Figures~2--4, and compared with
the adopted model spectrum in green.  It is seen that, while fitting the
He~I absorption lines well, the adopted model does not reproduce any of
the N~II or He~II emission lines; in fact, it does not reproduce the
N~III absorption-line strengths, either.  Fortunately, two other, unpublished 
trial models run at the same time are available; they are also shown here, 
in blue and red.  The blue model, with an increased N abundance (Table~1), 
does not produce emission lines, either.  However, the red model, with a
similarly increased N abundance, but also somewhat lower effective
temperature and increased mass-loss rate, does produce both N~II and
He~II emission lines; actually, the N~II emission is somewhat overestimated 
(Table~2).  On the other hand, the He~I lines are badly underproduced, 
which is why this model was not favored at the time.  D.~Lennon (priv. comm.) 
has suggested that increasing the He abundance above the number ratio to H of
0.2 adopted by Evans et~al.\ (2004) may remedy this discrepancy.  The results
discussed here indicate that it may well be possible to reproduce most
features in the optical spectrum of HDE~269896 with relatively small 
adjustments to the model parameters.

It is also very important to understand the atomic processes that produce
the selective emission lines in O-type spectra, as emphasized by Walborn
(2001).  These are lines that come into emission while others from the same 
ions remain in absorption; they are for the most part photospheric and 
respond to the temperature and luminosity.  A striking analogy to the
N~II lines discussed here, but at higher ionization, is provided by the 
selective emission in N~IV $\lambda$4058 vs. the absorption or P~Cygni 
profiles in the $\lambda$3480 blend from the same ion, in early-O spectra 
(Walborn et~al.\ 2004; Morrell et~al.\ 2005).  It is remarkable that current 
models can produce these emission lines; evidently the essential physics has
been correctly incorporated.  However, there has been no systematic effort 
to extract and elucidate the mechanisms involved, as was done for the Of N~III 
$\lambda\lambda$4634-4640-4642 triplet by Mihalas, Hummer, \& Conti (1972) 
and Mihalas \& Hummer (1973).  Such an effort will certainly provide 
interesting insights into the ionic level-population processes, and in turn 
further sensitive diagnostics for hot atmospheres.

Finally, it is interesting to compare HDE~269896 to other objects showing
these N~II emission lines, particularly the WN10 and WN11 types defined
by Crowther \& Smith (1997); see also Smith, Crowther, \& Prinja (1994).
These objects have much more extensive emission-line spectra, including a
strong He~I $\lambda$5016 P Cygni profile adjacent to and blended with
the N~II feature, which diagnose much denser winds.  Their typical He/H
ratios range from 0.3 to 0.6.  Clearly HDE~269896 is a highly evolved object, 
but less so than the WN10/11 stars.  Thus, it is reasonable to propose that 
it may be in an immediate pre-WNVL (Very Late WN) stage, i.e., that it will 
develop a denser, slower wind, a more extensive emission-line spectrum, and 
a higher surface He/H ratio as its evolution proceeds.  Many (perhaps all) 
WNVL objects are now recognized as quiescent states of Luminous Blue 
Variables (Walborn et~al.\ 2008 and references therein), and HDE~269896 has 
a comparable luminosity (Humphreys \& Davidson 1994).  HDE~269896 is 
located in relative isolation, north of 30~Doradus, but near the O2~III(f*) 
star Sk~$-68^{\circ}137$, also a very massive, likely WN progenitor 
(Walborn et~al.\ 2002b).

\acknowledgments
We thank the Director and staff of CASLEO for the use of their facilities. 
We also acknowledge the use at CASLEO of a CCD and data acquisition system 
partly financed by NSF Grant AST-90-15827 to Dr. R.M. Rich.  This research 
has received financial support from Instituto de Astrof\'{\i}sica La Plata 
(IALP), an institute of CONICET, Argentina.  Publication support was provided 
by the STScI Director's Discretionary Research Fund.  We thank Paul Crowther, 
the referee, for promoting an expanded scope of this paper.

\newpage

\begin{deluxetable}{llcccc}
\tablewidth{0pt}
\tablecaption{Model Fit Parameters}
\tablehead{
\colhead{Plot} &\colhead{Description} &\colhead{$T_\mathrm{eff}$} &\colhead{$\log(L/L_{\odot})$} 
&\colhead{$\log(\mathrm{N/H)}+12$} &\colhead{$\dot{M}$}\\
\colhead{Color} &&\colhead{(kK)} &&&\colhead{($M_{\odot}$ yr$^{-1}$)} 
}
\startdata
Green  &Evans et al.\ (2004) &27.5 &5.97 &8.3 &$7.5 \times 10^{-6}$\\
Blue    &Increased N           &27.0 &5.97 &8.9 &$7.5 \times 10^{-6}$\\
Red     &Increased N           &26.0 &5.97 &8.9 &$8.0 \times 10^{-6}$\\
          &Reduced $T_\mathrm{eff}$
\enddata
\end{deluxetable}

\begin{deluxetable}{cccrrccl}
\tabletypesize{\footnotesize}
\tablewidth{0pt}
\tablenum{2}
\tablecaption{N\,\textsc{ii} Line Parameters}
\tablehead{
\colhead{Wavelength} &\colhead{Rel.} &\colhead{Multiplet} &\multicolumn{4}{c}{Equivalent Widths (m\AA)} &\colhead{Comment}\\
\colhead{(\AA)}          &\colhead{Int.} &  &{Obs.}  &{Red} &\colhead{Green} &\colhead{Blue}
}
\startdata
\noalign{\vspace{-6pt}}
\multicolumn{6}{l}{Absorption or P Cygni Lines:}\\
\hline
\noalign{\vspace{2pt}}
3995.0      &1000    &$3p ^1\!D\rightarrow3s ^1\!P^0$  &$-13$ &\phn\phs50 &53 &\llap{1}78\\
4601.5      &\phn550    &$3p ^3\!P\rightarrow3s ^3\!P^0$   &$-8$\rlap{:} &\phn\phs38 &16 &82\\
4607.2      &\phn450    &" &\nodata &\nodata &\nodata &\nodata\\ 
4613.9      &\phn360    &"  &\nodata &\nodata &\nodata &\nodata\\ 
4621.4      &\phn450   &"  &\nodata &\nodata &\nodata &\nodata\\
4630.5      &\phn870   &"  &52 &\phn\phs96 &48 &\llap{1}75\\
4643.1      &\phn550   &"  &\nodata &\nodata &\nodata &\nodata &blended w.\ N\,\textsc{iii}\\
\hline
\multicolumn{6}{l}{Selective Emission Lines:}\\
\hline
\noalign{\vspace{2pt}}
4987.4      &\phn285   &$3d ^3\!P^0\rightarrow3p ^3\!S$ &$-16$ &\phn$-30$ &\nodata &\nodata\\
4994.4      &\phn450  &" &$-46$ &\phn$-68$ &\nodata &\nodata\\
5001.1      &\phn550  &$3d ^3\!F^0\rightarrow3p ^3\!D$ &$-120$ &$-180$ &13 &72 &blended\\
5001.5      &\phn650  &"   &\nodata &\nodata &\nodata &\nodata &blended\\
5002.7      &\phn360   &$3p ^3\!S\rightarrow3s ^3\!P^0$ &\nodata &\nodata &\nodata &\nodata &blended\\
5005.2      &\phn870  &$3d ^3\!F^0\rightarrow3p ^3\!D$ &$-65$ &$-240$ &\phn7 &50 &red model blend\\
5007.3      &\phn550  &$3d ^3\!P^0\rightarrow3p ^3\!S$ &$-68$ &\multicolumn{1}{c}{"} &\nodata &\nodata &red model blend\\
5010.6      &\phn450  &$3p ^3\!S\rightarrow3s ^3\!P^0$ &\nodata &\nodata  &\nodata &\nodata\\
5016.4      &\phn360    &$3d ^3\!F^0\rightarrow3p ^3\!D$ &\nodata &\nodata &\nodata &\nodata &blended w.\ He\,\textsc{i}\\
5025.7      &\phn360    &" &$-8$ &\phn$-13$ &\phn7 &18\\
\enddata
\end{deluxetable}

\begin{figure}
\includegraphics[width=\textwidth]{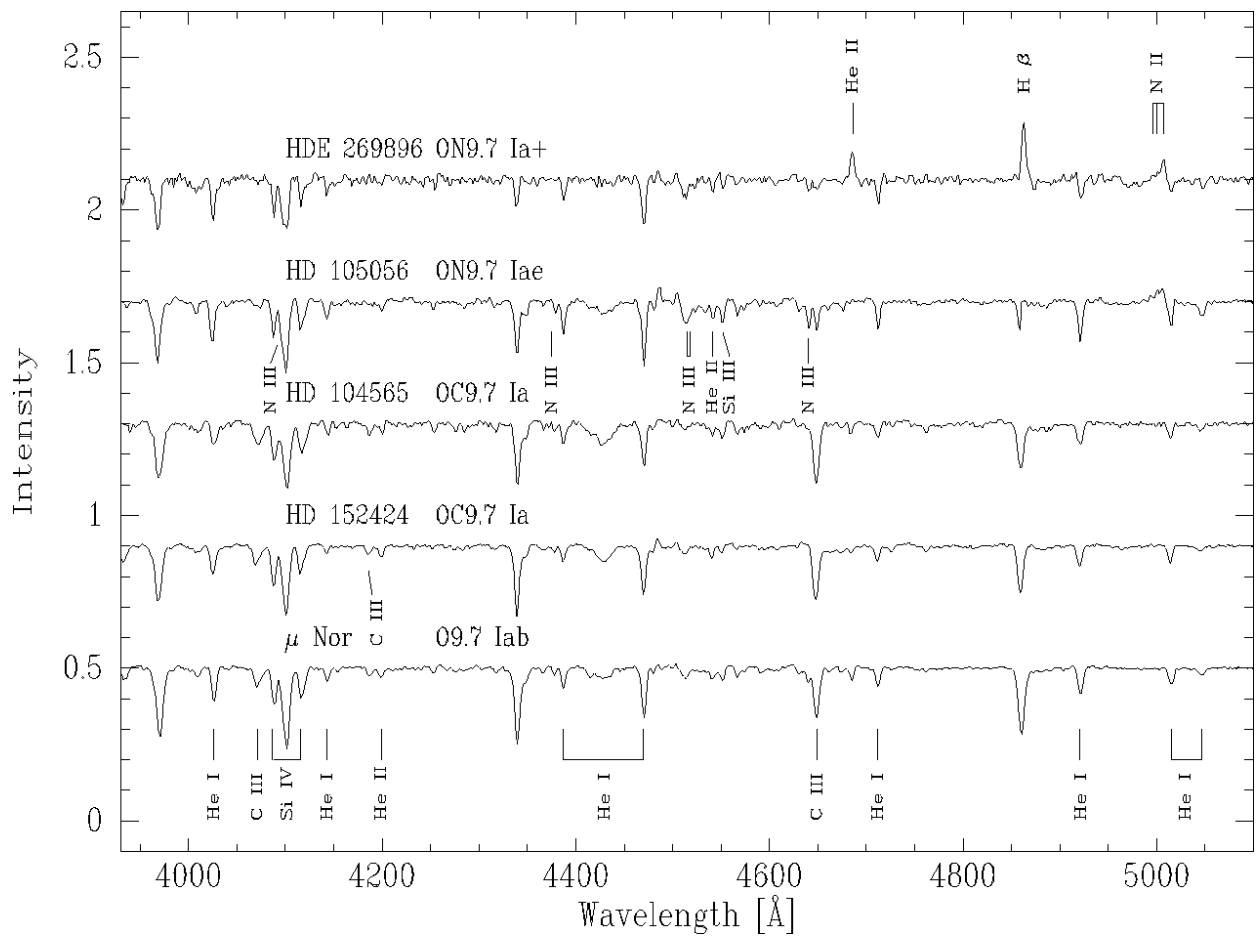}
\caption{\label{fig:fig1}
Violet through green digital, low-resolution spectrograms of HDE~269896 and 4 
comparison objects.  The intensity scale is rectified continuum units.  The
absorption lines identified below are He~I $\lambda\lambda$4026, 4144,
4387+4471, 4713, 4922, 5016+5048; C~III $\lambda\lambda$4070 and 4650
blends; Si~IV $\lambda\lambda$4089+4116; and He~II $\lambda$4200.  In the
spectrum of HD~152424, C~III $\lambda$4187 is identified; and in HD~105056, 
N~III $\lambda\lambda$4097 (blended with H$\delta$), 4379, 4511--4514 and 
4640-4642 blends; He~II $\lambda$4541; and Si~III $\lambda$4552.
The emission lines identified above are He~II $\lambda$4686, H$\beta$
$\lambda$4861, and the N~II $\lambda\lambda$4987--4994--5001--5005--5007
blend.}
\end{figure}

\begin{figure}
\includegraphics[width=\textwidth]{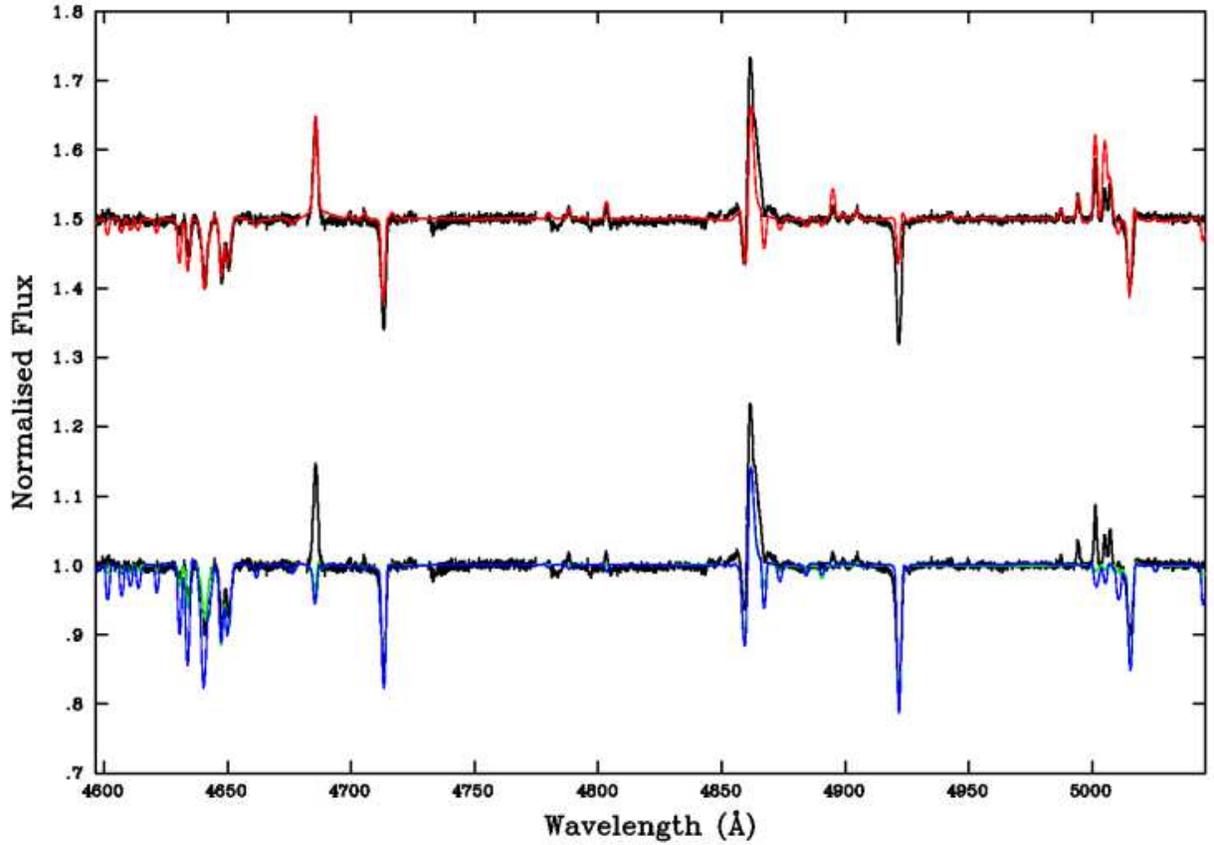}
\caption{\label{fig:fig2}
Green high-resolution spectrogram (black) and models (colors, Table~1) of
HDE~269896.  Rest wavelengths are shown.  The salient spectral features
are the N~II $\lambda\lambda$4600, 5000 complexes (Table~2); N~III
$\lambda$4640 and C~III $\lambda$4650 absorption blends; He~II
$\lambda$4686 emission; He~I $\lambda\lambda$4713, 4922, 5016
absorptions; and the H$\beta$ $\lambda$4861 P~Cyg profile.}
\end{figure}

\begin{figure}
\includegraphics[width=\textwidth]{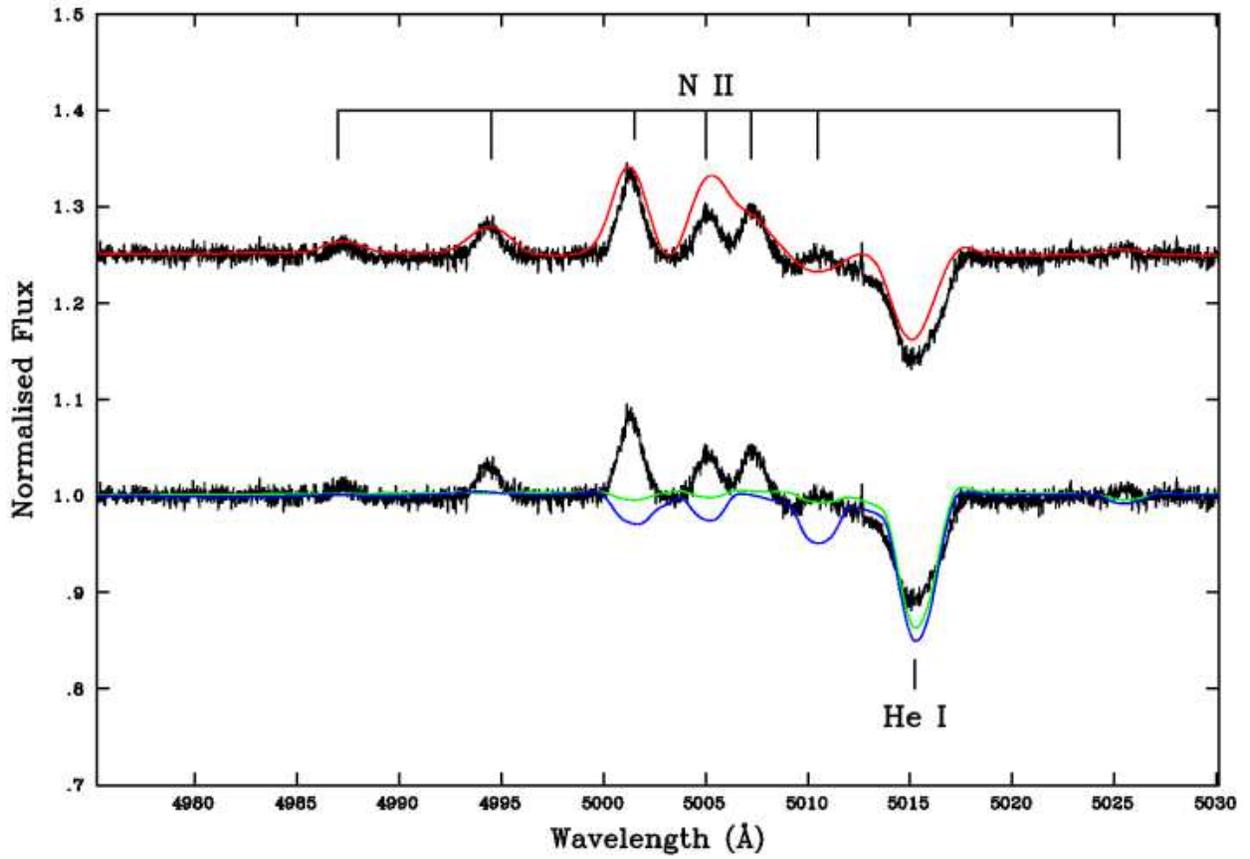}
\caption{\label{fig:fig3}
Enlargement of the $\lambda$5000 region in Fig.~1.  Wavelengths of the
N~II lines are given in Table~2.}
\end{figure}

\begin{figure}
\includegraphics[width=\textwidth]{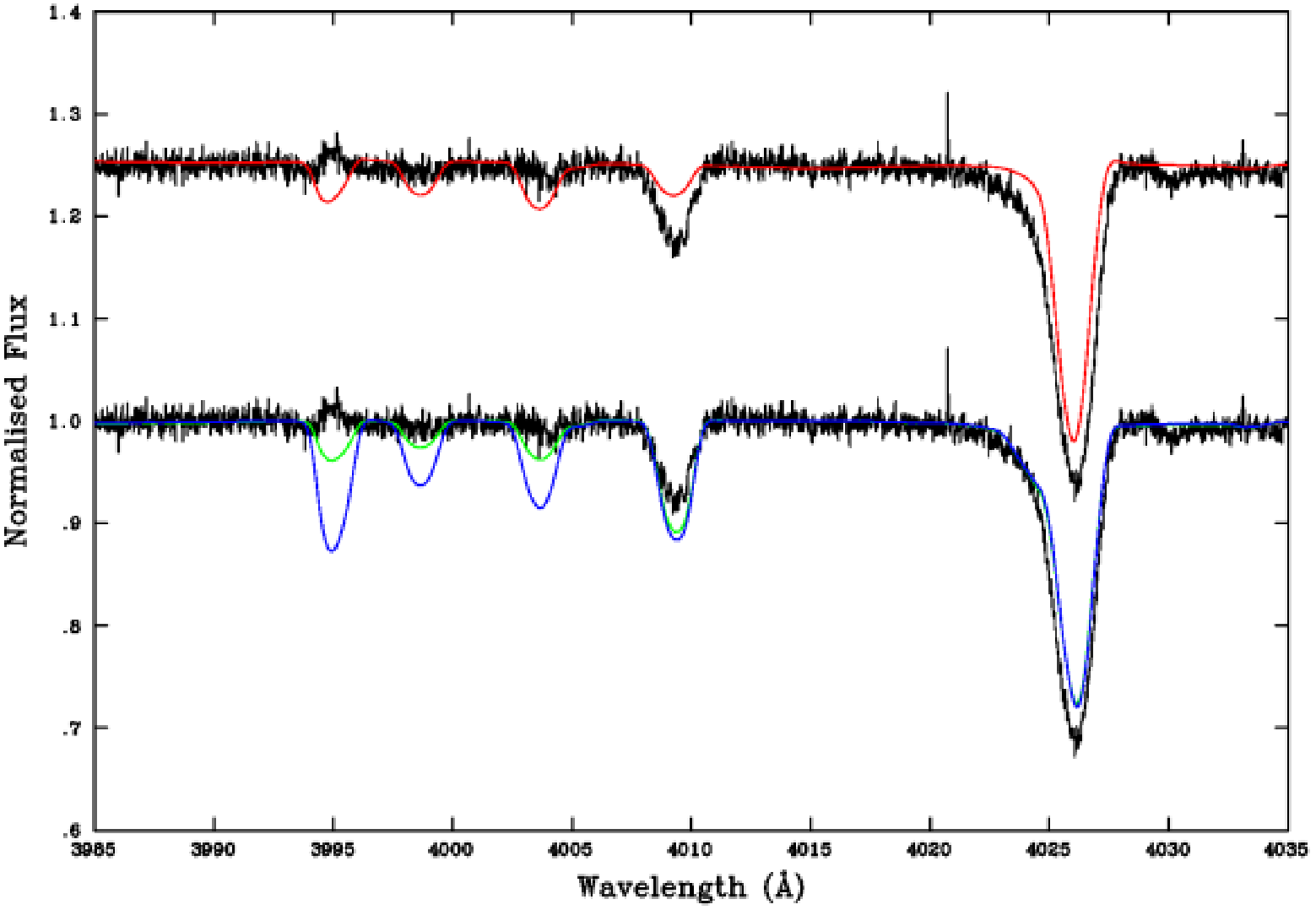}
\caption{\label{fig:fig4}
Enlargement of the $\lambda$4000 region in the high-resolution
spectrogram of HDE~269896.  The spectral lines are N~II $\lambda$3995;
N~III $\lambda\lambda$3999, 4004; He~I $\lambda$4009; and He~I+II
$\lambda$4026.}
\end{figure}


\begin{references}
\reference{} Crowther, P.A., Hillier, D.J., Evans, C.J., Fullerton, A.W.,
   De Marco, O., \& Willis, A.J. 2002, \apj, 579, 774 
\reference{} Crowther, P.A., \& Smith, L.J. 1997, \aap, 320, 500
\reference{} Evans, C.J., Crowther, P.A., Fullerton, A.W., \& Hillier, D.J.
   2004, \apj, 610, 1021
\reference{} Fitzpatrick, E.L. 1991, \pasp, 103, 1123
\reference{} Humphreys, R.M., \& Davidson, K. 1994, \pasp, 106, 1025
\reference{} Hyland, A.R., \& Bessell, M.S. 1975, Proc. Astr. Soc.
   Australia, 2, 353
\reference{} Maeder, A., \& Meynet, G. 2000, \araa, 38, 143
\reference{} Mihalas, D., \& Hummer, D.G. 1973, \apj, 179, 827
\reference{} Mihalas, D., Hummer, D.G., \& Conti, P.S. 1972, \apj, 175, L99
\reference{} Morrell, N.I., Walborn, N.R., \& Arias, J.I. 2005, \pasp, 
   117, 699
\reference{} Ralchenko, Yu., Kramida, A.E., Reader, J., \& NIST ASD Team 2008, 
   NIST Atomic Spectra Database (v3.1.5), http://physics.nist.gov/asd3   
\reference{} Smith, L.J., Crowther, P.A., \& Prinja, R.K. 1994, \aap, 281, 833
\reference{} Walborn, N.R. 1971, \apjs, 23, 257
\reference{} ----- 1973, \aj, 78, 1067
\reference{} ----- 1976, \apj, 205, 419 
\reference{} ----- 1977, \apj, 215, 53
\reference{} ----- 2001, in ASP Conf. Ser., 242, Eta Carinae \& Other Mysterious
   Stars, ed. T. Gull, S. Johansson, \& K. Davidson (San Francisco: ASP), 217
\reference{} ----- 2003, in ASP Conf. Ser., 304, CNO in the Universe,
   ed. C. Charbonnel, D. Schaerer, \& G. Meynet (San Francisco: ASP), 29   
\reference{} ----- 2008, in Spectral Classification, ed. R.O. Gray \&
   C. Corbally (Princeton University Press), in press
\reference{} Walborn, N.R., \& Fitzpatrick, E.L. 1990, \pasp, 102, 379
\reference{} Walborn, N.R., Fullerton, A.W., Crowther, P.A., Bianchi, L.,
   Hutchings, J.B., Pellerin, A., Sonneborn, G., \& Willis, A.J. 2002a,
   \apjs, 141, 443
\reference{} Walborn, N.R., \& Howarth, I.D. 2000, \pasp, 112, 1446
\reference{} Walborn, N.R., Howarth, I.D., Lennon, D.J., Massey, P.,
   Oey, M.S., Moffat, A.F.J., Skalkowski, G., Morrell, N.I., Drissen, L.,
   \& Parker, J.Wm. 2002b, \aj, 123, 2754
\reference{} Walborn, N.R., Morrell, N.I., Howarth, I.D., Crowther, P.A.,
   Lennon, D.J., Massey, P., \& Arias, J.I. 2004, \apj, 608, 1028 
\reference{} Walborn, N.R., Parker, J.Wm., \& Nichols, J.S. 1995,
   International Ultraviolet Explorer Atlas of B-Type Spectra from
   1200 to 1900 \AA, NASA RP 1363
\reference{} Walborn, N.R., Stahl, O., Gamen, R.C., Szeifert, T.,
   Morrell, N.I., Smith, N., Howarth, I.D., Humphreys, R.M., Bond, H.E., 
   \& Lennon, D.J. 2008, \apj, 683, L33
\end{references}
\end{document}